\documentstyle[12pt]{article}
\catcode`\@=11

\catcode`\@=11

\def\section{\@startsection{section}{1}{\z@}{3.5ex plus 1ex minus
   .2ex}{2.3ex plus .2ex}{\large\bf}}

\def\ps@headings{\def\@oddfoot{}\def\@evenfoot{}
\def\@oddhead{\hbox{}\hfill
        \makebox[.5\textwidth]{\raggedright\ignorespaces --\thepage{}--
        \hfill }}
\def\@evenhead{\@oddhead}
\def\subsectionmark##1{\markboth{##1}{}}
}

\ps@headings

\catcode`\@=12

\relax

\def\figcap{\section*{Figure Captions\markboth
        {FIGURECAPTIONS}{FIGURECAPTIONS}}\list
        {Fig. \arabic{enumi}:\hfill}{\settowidth\labelwidth{Fig. 999:}
        \leftmargin\labelwidth
        \advance\leftmargin\labelsep\usecounter{enumi}}}
 \relax
\def\tablecap{\section*{Table Captions\markboth
        {TABLECAPTIONS}{TABLECAPTIONS}}\list
        {Table \arabic{enumi}:\hfill}{\settowidth\labelwidth{Table 999:}
        \leftmargin\labelwidth
        \advance\leftmargin\labelsep\usecounter{enumi}}}
 \relax
\def\reflist{\section*{References\markboth
        {REFLIST}{REFLIST}}\list
        {[\arabic{enumi}]\hfill}{\settowidth\labelwidth{[999]}
        \leftmargin\labelwidth
        \advance\leftmargin\labelsep\usecounter{enumi}}}
 \relax

\catcode`\@=11

\def\marginnote#1{}
\newcount\hour
\newcount\minute
\newtoks\amorpm
\hour=\time\divide\hour by60
\minute=\time{\multiply\hour by60 \global\advance\minute by-
\hour}
\edef\standardtime{{\ifnum\hour<12 \global\amorpm={am}%
    \else\global\amorpm={pm}\advance\hour by-12 \fi
    \ifnum\hour=0 \hour=12 \fi
    \number\hour:\ifnum\minute<100\fi\number\minute\the\amorpm}}
\edef\militarytime{\number\hour:\ifnum\minute<100\fi\number\minute}
\def\draftlabel#1{{\@bsphack\if@filesw {\let\thepage\relax
  \xdef\@gtempa{\write\@auxout{\string
    \newlabel{#1}{{\@currentlabel}{\thepage}}}}}\@gtempa
    \if@nobreak \ifvmode\nobreak\fi\fi\fi\@esphack}
     \gdef\@eqnlabel{#1}}
\def\@eqnlabel{}
\def\@vacuum{}
\def\draftmarginnote#1{\marginpar{\raggedright\scriptsize\tt#1}}
\def\draft{\oddsidemargin -.5truein
        \def\@oddfoot{\sl preliminary draft \hfil
        \rm\thepage\hfil\sl\today\quad\militarytime}
        \let\@evenfoot\@oddfoot \overfullrule 3pt
        \let\label=\draftlabel
        \let\marginnote=\draftmarginnote
   
\def\@eqnnum{(\theequation)\rlap{\kern\marginparsep\tt\@eqnlabel}%
\global\let\@eqnlabel\@vacuum}  }
\def\preprint{\twocolumn\sloppy\flushbottom\parindent 1em
        \leftmargini 2em\leftmarginv .5em\leftmarginvi .5em
        \oddsidemargin -.5in    \evensidemargin -.5in
        \columnsep 15mm \footheight 0pt
        \textwidth 250mmin      \topmargin  -.4in
        \headheight 12pt \topskip .4in
        \textheight 175mm
        \footskip 0pt
        
\def\@oddhead{\thepage\hfil\addtocounter{page}{1}\thepage}
        \let\@evenhead\@oddhead \def\@oddfoot{} \def\@evenfoot{} 
}
\def\titlepage{\@restonecolfalse\if@twocolumn\@restonecoltrue\onecolumn
     \else \newpage \fi \thispagestyle{empty}\c@page\z@
        \def\thefootnote{\fnsymbol{footnote}} }
\def\endtitlepage{\if@restonecol\twocolumn \else  \fi
        \def\thefootnote{\arabic{footnote}}
        \setcounter{footnote}{0}}  %\c@footnote\z@ }
\catcode`@=12
\relax
\def\ps@headings{\def\@oddfoot{}\def\@evenfoot{}
\def\@oddhead{\hbox{}\hfill
        \makebox[.5\textwidth]{\raggedright\ignorespaces --\thepage{}--
        \hfill }}
\def\@evenhead{\@oddhead}
\def\subsectionmark##1{\markboth{##1}{}}
}

\ps@headings

\relax

\def\firstpage#1#2#3#4#5#6{
\begin{document}
%\draft
%\input epsf.tex
\newcommand{\newc}{\newcommand} 
\newc{\ra}{\rightarrow} 
\newc{\lra}{\leftrightarrow} 
\newc{\beq}{\begin{equation}} 
\newc{\eeq}{\end{equation}} 
\newc{\bea}{\begin{eqnarray}} 
\newc{\eea}{\end{eqnarray}}
\begin{titlepage}
\nopagebreak
\title{\begin{flushright}
        \vspace*{-0.8in}
\end{flushright}
\vfill
{#3}}
\author{\large #4 \\[1.0cm] #5}
\maketitle
\vskip -7mm     
\nopagebreak 
\begin{abstract}
{\noindent #6}
\end{abstract}
\vfill
\begin{flushleft}
\rule{16.1cm}{0.2mm}\\[-3mm]
$^{\star}${\small Research supported in part by the EEC under the 
\vspace{-4mm} TMR contract ERBFMRX-CT96-0090 and in part by the
project CSI-430C.}\\ 
September 1997
\end{flushleft}
\thispagestyle{empty}
\end{titlepage}}

\def\simlt{\stackrel{<}{{}_\sim}}
\def\simgt{\stackrel{>}{{}_\sim}}
\date{}
\firstpage{3118}{IC/95/34}
{\large\bf   
Octonionic  Selfduality for SuperMembranes$^{\star}$} 
{E.G. Floratos$^{\,a,b}$ and G.K. Leontaris$^{\,c,d}$}%\\[-3mm] 
{\normalsize\sl
$^a$Institute of Nuclear Physics,  NRCS Demokritos,
{} Athens, Greece\\[-3mm]
\normalsize\sl
$^b$ Physics Deptartment, University of Iraklion, 
Crete, Greece.\\[-3mm]
\normalsize\sl
$^c$Theoretical Physics Division, Ioannina University,
GR-45110 Ioannina, Greece.\\[-3mm]
\normalsize\sl
$^d$ CERN, Theory Division, 1211 Geneva, 23, Switzerland.}
{In this work we study the recently introduced octonionic duality for membranes.
Restricting the self - duality  equations to seven space dimensions, we provide
various forms for them  which exhibit the  symmetries of the octonionic and 
quaternionic structure. These forms may turn to be useful for the  question 
of the integrability of this system. Introducing a consistent quadratic
Poisson algebra of functions on the membrane  we are able to factorize the
time dependence of the self - duality equations. We further give the general 
linear  embeddings of the three dimensional system into the seven dimensional one
using the invariance of the self-duality equations under the exceptional group $G_2$.
}
\newpage
\newpage
\section{Introduction.}
M theory is the leading candidate for the unification of all superstring theories in their
perturbative and non-perturbative sector. This theory contains
$N=1$, 11-dimensional supergravity and at least a sector of 
supermembranes and their magnetic duals the superfive~branes\cite{EW0,rev}.
 These extended objects exist as 
solitons of the eleven dimensional supergravity and they are distinguished 
from the fundamental superbranes as solitonic superbranes.\cite{xx}

Most of the recent work on  compactifications of M  theory is concentrated on a unified ``proof''
of various non-perturbative dualities of superstring theories connecting their strong and weak 
coupling sectors or small with large volumes of the compactifying manifolds\cite{EW0,rev}.
There is a line of attack from the point of view of the 11-d superbranes which either uses
double dimensional reduction to connect with type IIA or heterotic superstrings, or using
purely classical worldvolume dualities of the superbranes which happens miraculously to explain
non - perturbative phenomena, (dualities of superstring theories)\cite{revtownsend}. 
 {}From the point of view of superstring theories, 
supermembranes in non-compact 11 dimensions correspond to the strong coupling regime of the
superstrings\cite{EW0}.

Many basic questions concerning supermembrane theories have not been answered today.
A top priority issue, is  the derivation of the eleven dimensional supergravity theory as a low energy 
effective action of the supermembrane. To do this one needs to understand the quantum mechanics of
the supermembrane, that is, define a sensible perturbation theory. This is  an extremely
hard problem for two reasons. First, the moduli space of three dimensional Remannian 
metrics is largely unknown and a representation theory of the three dimensional diffeomorphism
group or ( in the light-cone gauge ) of the area preserving diffeomorphism group of the 
supermembrane is lacking. Second, -- unlike the string where in the light
cone gauge the theory becomes an infinite set of free  transverse oscillators --  
in the supermembrane case,  the light cone gauge does not  solve fully the constraints and
there is no coupling constant in the interaction term.
 {}Fortunately, the Hamiltonian in
the light cone gauge is of Yang-Mills (YM) type  with gauge group the area preserving diffeomorphisms
of the membrane\cite{HpTh,Hope,BST}.

Another issue is the following. During the compactifications from eleven dimensions to ten, one
is freezing an infinite number of string degrees of freedom of the supermembrane and considers
only the Kaluza-Klein  dilatonic modes which are supposed to be the infinite tower of superstring
solitons which complete the duality picture. It could be possible that taking into account
in a controlable way the interaction of the remaining string excitations of the supermembrane,
one could define a perturbation theory\cite{RZ}.
 Recently, an old mode regularization of supermembrane
through $SU(N)$ matrix super YM mechanics has been re-incarnated as possible candidate model for
M theory\cite{Banks1}

Another possible approach to define a perturbative expansion for the 11 dimensional supermembrane
is to study various compactifications of the 11-d supergravity where the classical supermembrane 
has very simple dynamics (it can be even static-stretched) and then around these classical 
solutions study the quantum excitations of the supermembrane.  In this way one hopes to get
a classical state which could be used as a quantum vacuum state for the membrane. One test
would be to find in the excitation spectrum of the supermembrane the 11-d supergravity multiplet 
around the classical background. The problem is that one has to preserve in one 
way or another the $N=1$,
11-d sypersymmetry during these compactifications\cite{xx,revtownsend}.
{}Following old work in the compactification of 11-d supergravity on the seven sphere\cite{old}
there is a recent activity on octonionic solitons for strings and supermembranes\cite{Duff}.
In this work, specific background field configurations of the compactified supergravity on
seven sphere considered as various fiberbundles which  are coupled through their singularities
to supermembrane sources.

In this work, we want to move in different direction which exploits some aspects of the non
perturbative structure of the supermembrane vacuum in flat space time, studying classical
Euclidean time equations  which describe quantum tunneling processes
between  classical configurations  of the supermembrane which could be considered as vacua 
of different topological sectors. Although there is an extensive work 
\cite{xx}, where essentially the background field equations are solved,
 as far as we know, the question of the Euclidean membrane as an extended object
connecting different topological sectors has not been addressed except in \cite{Sav,FL,FCZ}.
The topological charge and the Bogomol'nyi bound known from supersymmetric YM theory,
can be extended to Euclidean supermembranes in $(4+1)$\cite{Sav,FL}
and, as has been shown recently, in $(8+1)$-dimensions\cite{FCZ}.
In section 2 we recall the main results of the works\cite{Sav,FL} where the self dual bosonic
membrane in $(2+1)$ and (4+1) dimensions 
 has been introduced. In section 3, the generalization by\cite{FCZ} in (8+1) dimensions
is described in a compact form and possible factorizations of the time dependence are discussed.
{}In section 4, the same equations in octonionic and quartenionic representations
are introduced which exhibit specific properties of the self-duality equations. Finally,
is section 5 the general formulation of embedding the three
 dimensional solutions into seven dimensions 
is described and the constraint equations are derived.  Some examples of specific embeddings of
 the (4+1)-dimensional system 
to $(8+1)$ dimensions are also analysed.
%%%%%%%%%%%%%%%%%%%%%%%%%%%%%%%%%%%%%%%%%%%%%%%%
\section{SU(N) Yang Mills and Membranes.}
To start,  we recall that it has been known
since sometime that the supermembrane Hamiltonian in the light-cone gauge is a very close
relative of Yang-Mills (YM) theories in the gauge $A_0=0$ and in one space dimension less
\cite{HpTh,Hope}.
 To describe in some more detail this relationship, we restrict our discussion to the
bosonic part of the Hamiltonian of the supermembrane in the light cone gauge
 and to spherical topology for the membrane\cite{Hope,FIT,FL}. 
In reference\cite{FIT} using results of[5],  it was pointed out that in the large $N$-limit,
$SU(N)$ YM theories have, at the classical level, a simple
geometrical structure with the SU(N) matrix potentials $A_{\mu}(X)$ replaced by
c-number functions of two additional coordinates $\theta, \phi$ of an internal
sphere $S^2$ at every space-time point,
 while the $SU(N)$ symmetry is replaced by
the infinite dimensional algebra of area preserving diffeomorhisms
 of the sphere
$S^2$ called SDiff$(S^2)$. 
The $SU(N)$ fields ($N\times N$ matrices)
\begin{eqnarray}
& &A_{\mu}(X)=A^{\alpha}_{\mu}(X)t^{\alpha},\nonumber\\
& &t^{\alpha}\in SU(N),\nonumber\\
& &\alpha=1,2,\ldots,N^2-1,\quad \mu=0,1,\dots d-1 \label{1}
\end{eqnarray}
in the large $N$ - limit become $c$-number functions of an
internal sphere $S^2$,
\begin{equation}
A_{\mu}(X,\theta,\phi)=\sum^{\infty}_{l=1}\sum^l_{m=-l}
A^{lm}_{\mu}(X)Y_{lm}(\theta,\phi),\label{2}
\end{equation}
where $Y_{lm}(\theta,\phi)$ are the spherical harmonics on $S^2$.
 The local gauge
transformations
\begin{equation}
\delta A_{\mu}=\partial_{\mu}\omega+[A_{\mu},\omega],\quad 
\omega=\omega^{\alpha}t^{\alpha}, \label{3a}
\end{equation}
and
\begin{equation}
\delta F_{\mu\nu}=[ F_{\mu\nu},\omega],\label{3b}
\end{equation}
\begin{equation}
F_{\mu\nu}=\partial_{\mu}A_{\nu}-\partial_{\nu}A_{\mu}+
[A_{\mu},A_{\nu}] \label{lgt}
\end{equation}
are replaced by
\begin{equation}
\delta A_{\mu}(X,\theta,\phi)=\partial_{\mu}\omega
(X,\theta,\phi)+\{A_{\mu},\omega\},\label{4a}
\end{equation}
\begin{equation}
\delta F_{\mu\nu}(X,\theta,\phi)=\{F_{\mu\nu},\omega\},
\end{equation}
where
\begin{equation}
 F_{\mu\nu}(X,\theta,\phi)=\partial_{\mu}A_{\nu}-
\partial_{\nu}A_{\mu}+\{A_{\mu},A_{\mu}\},\label{5} 
\end{equation}
and the Poisson bracket on $S^2$ is defined as follows:
\begin{equation}
\{f,g\}=\frac{\partial f}{\partial\phi}\frac{\partial g}
{\partial\cos\theta}-\frac{\partial g}{\partial\phi}
\frac{\partial f}{\partial\cos\theta} \label{6} 
\end{equation}
So the commutators are replaced by Poisson brackets according to
\begin{equation}
\lim_{N\rightarrow\infty}N[A_{\mu},A_{\nu}]=\{A_{\mu},A_{\nu}\}
\label{7}
\end{equation}
Then the YM action in the large-$N$ limit becomes \cite{FIT}
\begin{equation}
S_{\infty}=\frac{1}{16\pi g^2}\int_{S^2}d\Omega\int d^4XF_{\mu\nu}
(X,\theta,\phi)F^{\mu\nu}(X,\theta,\phi), \label{8}
\end{equation}
where 
\begin{equation}
g=\lim_{N\rightarrow\infty}\frac{g_N}{N^{3/2}} \label{9}
\end{equation}
This large-$N$ limit of $SU(N)$ YM theories was found by making use of the
relation between the SU(N=2 s+1) algebra in a particular basis 
(up to spin $s$ SU(2)- tensor $N\times N$- matrices) and  $SDiff(S^2)$ 
in the basis of the spherical harmonics $Y_{lm}(\theta,\phi)$.
In the present day language, this  $SDiff(S^2)$ YM theory corresponds to the
effective theory of infinite number, 
$N\ra \infty$, $d-1$ dimensional Dirichlet branes\cite{JP,PT1}.
Similar considerations hold for membranes of different topologies, torus,
 double torus etc\cite{Bars}. 
Here we note that the recently proposed matrix theory
which is claimed to be the long seeked formulation of M theory, is nothing
but the SU(N) supersymmetric YM mechanics which was used as a consistent 
truncation of the supermembrane\cite{HpTh,Hope,Banks}.

The above considered large $N$-limit, is a very specific one which depends on
the appropriate basis of $SU(N)$ generators convenient for the topology of
the membrane and it has nothing to do, at least in a direct way, with the
planar approximation of YM theories. Also, it is different from the large N-limit 
used in Matrix theory.

In the case of the spherical membranes the  $SDiff(S^2)$ YM- theory describes the dynamics
of an infinite number of $D0$- branes forming a topological 2- sphere. In the
light cone gauge the transverse coordinates $X_i$, ($i,1...,9$) of the 
11-d bosonic part of the supermembrane satisfy the
following equations
\beq
\ddot{X}_i = \{X_k,\{X_k,X_i\}\}
\;\; i,k = 1,\ldots 9\label{eom}
\eeq
where summation over repeated indices is implied. The corresponding Gauss law which is the
generator of the  $SDiff(S^2)$ group is given by the constraint
\beq
\{X_i,\dot{X}_i\} = 0\label{GL}
\eeq
In ref\cite{Sav} Euclidean bosonic membranes in 3 Dimensional target space  have
been introduced defining the topological charge density to be 
\beq
\Omega(X) = \frac{1}{3!}\epsilon^{abc}f_{ijk}X_a^iX_b^jX_c^k
\eeq
where, $a,b,c$ run from 1 to 3 and $i,j,k$ from 1 to $d$ space time dimensions.
\beq
X_a^i=\partial_{\xi_a}X^i
\eeq
and $\xi_{1,2,3}$ are the worldvolume coordinates.
The self-duality equations were defined as
\beq
P_i^a = \pm \frac 12\epsilon^{abc}f_{ijk}X_b^jX_c^k
\eeq
Here $P_i^a $ are the canonical momenta, 
\beq
P_i^a = T\frac{\delta}{\delta X_a^i}(Det[X_a^iX_b^i]))^{1/2}
\eeq
The self duality equations for the case $d=3$ and $f_{ijk}=\epsilon_{ijk}$ were shown
to satisfy both the constraints and the equations of motion. Solutions were given
for the case of sphere and torus. 
In reference\cite{FL} 3-d  Euclidean self duality  equations in the light cone gauge
(that is 4+1-dimensional target space) for the  bosonic
part of the supermembrane  could be written in analogy with the 3-d Nahm equations
of self dual BPS YM- monopoles. In the light cone gauge this means that one
had to fix 6 of the 9 transverse coordinates to be constants. This constraint
solves the second order equations (\ref{eom}) for the 6 coordinates. 
{}Then the self duality equations are
\beq
\dot{X}_i = \frac 12\epsilon_{ijk}\{X_j,X_k\}
\;\; i,j=1,2,3
\label{sde}
\eeq
The self duality equations solve automatically the second order Euclidean time equations
as well as the Gauss law due to the Jacobi identity for the $\epsilon$ symbol and its
well known properties. 
The above system has a Lax pair and infinite number of conservation laws\cite{FL}.
In order to see this, first we rewrite eqs.$(\ref{sde})$ in the form  
\begin{equation}
\dot{X}_+=i\{X_3,X_+\}, \;\;
\dot{X}_-=i\{X_3,X_-\}, \;\;
\dot{X}_3=\frac{1}{2}i\{X_+,X_-\}, \label{17c}
\end{equation}
where
\begin{equation}
X_{\pm}=X_1\pm iX_2 \label{18}
\end{equation}
 
There exists a linear system corresponding to $(\ref{17c})$ which is the following
\beq
\dot{\psi}=L_{X_3+\lambda X_-}\psi, \quad \dot{\psi}=L_{\frac{1}{\lambda}
X_{+}-X_3}\psi,\label{21a,b}
\eeq
where the differential operators $L_f$ are defined as
\beq
L_f\equiv i\big(\frac{\partial f}{\partial\phi}
\frac{\partial}{\partial\cos\theta}-\frac{\partial f}
{\partial\cos\theta}\frac{\partial}{\partial\phi}\big).\label{22}
\eeq
The compatibility condition of $(\ref{21a,b})$ is
\beq
[\partial_t-L_{X_3+\lambda X_-},\partial_t-L_{\frac{1}{\lambda}
X_{+}-X_3}]=0,\label{23}
\eeq
from which, comparing the two sides for the coefficients of the powers
$\frac{1}{\lambda},\lambda^0,\lambda^1$ of the spectral parameter $\lambda$, we
find $(\ref{17c})$. From the linear
system  $(\ref{21a,b})$ using the inverse scattering method, one could
in principle construct all solutions of the self-duality equations.

Specific solutions could be obtained due to the existence of an $SU(2)$ subalgebra
of $SDiff(S^2)$ which happens to be its only finite dimensional subalgebra.
 Using this $SU(2)$ subalgebra, for spherically symmetric
solutions it can be shown that the system reduces to the Toda $SU(2)$ equations.
Another method to find solutions of the integrable system (\ref{17c})
 has been proposed in\cite{Ward} where the system is linearized
by considering the target space variables as worldvolume variables and vice-versa.
More recently, there have been discussions of the same issue in papers\cite{FCZ,Za}.
In ref\cite{Pleb} the connection with self dual Einstein equations has been discussed.
Before closing the section, we would like to note that the Euclidean membrane configurations
which are solutions of the selfduality equations are expected to interpolate between
classical vacuum configurations of the membrane that is, points or strings.
Also the case of membrane is the first in the series of extended objects which there is a gauge
principle to define the interactions and the possibility arises for topology change through
gauge interactions. The case of string has an ad hoc interaction which is not enforced
uniquely by any gauge principle. Moreover, the classical vacua of string are points.\cite{Sav}

\section{The octonionic structure of the self duality equations}

An obvious way to generalize duality for super p-branes, is to use Poincar\'e 
duality. {}For the fundamental supermembranes in particular, this has been done by
Duff et al\cite{Dea} and it has been exploited later,  proving various 
conjectures of string-string, string-membrane and membrane-membrane dualities\cite{DLM,BBS,HT}.
Another type of duality has been investigated recently\cite{FCZ,F1} which is based on
the existence of the last real division algebra, the octonionic or Cauley algebra\cite{gursey}.
The  work of\cite{FCZ} is based on the similarity between the supermembrane and the super YM theories
refered previously and the work on 8-dimensional YM instantons many years  ago
by\cite{fairlie}. Another way of considering the work of ref\cite{FCZ} is as 
 extension of the quaternionic case\cite{FL}  using the possibility
to define a cross product of  two vectors in 8 dimensions through the  multiplication rule
of octonions.

In this section, we restrict the self duality equations of \cite{FCZ} to seven dimensions
by choosing fixed values for 8 and 9 membrane coordinates.
Then, the the self duality equations\cite{FCZ}  become  
\begin{equation}
\dot{X}_i = \frac{1}{2} \Psi_{ijk}\{X_j,X_k\}
\label{osce}
\end{equation}
where  $\Psi_{ijk}$ is the completely antisymmetric tensor which defines
the multiplications of octonions\cite{gursey}.
The Gauss law results automatically by making use of the $\Psi_{ijk}$ cyclic symmetry
\beq
\{\dot{X}_i,X_i\}= 0
\eeq
The Euclidean equations of motion  are obtained as follows
\bea
\ddot{X}_i& = & \frac 12\Psi_{ijk}\left(\{\dot{X}_j,X_k\}+ \{X_j,\dot{X}_k\}\right)\\
          & = & \{X_k,\{X_i,X_k\}\}
\eea
where use has been made of the identity
\beq
\Psi_{ijk}\Psi_{lmk}=\delta_{il}\delta_{jm}-\delta_{im}\delta_{jl}+\phi_{ijlm}
\label{I1}
\eeq
and  of the cyclic property of the symbol 
$\phi_{ijlm}$\cite{gursey}.

As in the case of the 3-d system we may try to factorize the time dependence. We assume the
following factorization.
\beq
X_i = Z_{ij}(t) f_j(\xi )
\eeq
Then, from Eq.(\ref{osce}) we obtain
\bea
\dot{Z}_{im}f_m& = & \frac 12\Psi_{ijk}Z_{jl}Z_{kn}\{f_l,f_n\}
\eea
We observe that if we  make the Ansatz for the $7\times 7$ matrix
\beq
\dot{Z}_{im}(t) \Psi_{mln} = \Psi_{ijk} Z_{jl}(t)Z_{kn}(t)
\eeq
then the equation
\beq
f_i = \frac{1}{2}\Psi_{ijk}\{f_j,f_k\}
\label{feq}
\eeq
is automatically satisfied, while at the same time
we have succeeded to disentangle the time dependence from the self-duality equation.
Therefore, the problem is reduced to find solutions for $f_i(\xi)$ and $Z_{kl}$ equations
seperately. 

Another equivalent form of the previous equation for
the matrices $Z_{ij}$ is 
\bea
\dot{Z}_{ij} &=& \frac 16\Psi_{ikl}\Psi_{jmn}Z_{km}Z_{ln}
\eea
In the case of diagonal matrices $Z_{ij} = \delta_{ij} R_j(t)$,
we have 
\bea
\dot{R}_i& =& \frac 16\Psi_{ikl}^2 R_k R_l
\eea
We make now some observations about  the symmetries
of Eqs(\ref{osce},\ref{feq}). If $X_i$ is a solution of (\ref{osce}) then
for every matrix $R$ of the group $G_2$ which is a subgroup of $SO(7)$ then
\beq
Y_i = R_{ij} X_j\label{Rij}
\eeq
is automatically a solution of the same equation because the elements of $G_2$
preserve the structure constants $\Psi_{ijk}$. In components,
\beq
\Psi_{ijk} R_{kl}= \Psi_{imn}R_{mj}R_{nl}
\label{orth}
\eeq
The above relation shows the  way to define $G_2$ group elements
starting from two orthonormal 7-vectors.
The equation is obviously covariant under $SDiff(S2)$ transformations.
One can define combined $G_2$ and $SDiff(S2)$ transformations to get
 $SO(3)$ spherically symmetric solutions since $SO(3)$ can be realised
as a subalgebra of $SDiff(S2)$.

We note that in principle it is possible to look for non-linear symmetries
of the self-duality  equations, generalizing (\ref{Rij}) 
\beq
Y_i = f_i(X)
\eeq
where $f_i(X)$ must satisfy the equation
\beq
\Psi_{ijk}\frac{\partial f_k}{\partial X_l}= \Psi_{imn}\frac{\partial f_m}{\partial X_j}
\frac{\partial f_n}{\partial X_l}
\eeq

 In the following we examine  the self consistency of Eq.(\ref{feq}).
 Multiplying by $\Psi_{ilm}$, we get
\beq
\Psi_{ilm} f_i = \{f_l,f_m\}+\frac 12\phi_{lmjk}\{f_j,f_k\}
\eeq
Then, since the Poisson brackets satisfy the Jacobi identity, the above equation
 is constrained to satisfy the identity
\beq
\frac 13\phi_{ijkl}f_l =\Psi_{ijn}\{f_m,f_k\}+ {\rm cyclic\; perm.\; of} (ijk)
\eeq
This system of  equations  is exactly the same as in (\ref{feq}).

Another check for the self consistency of $f_i$ equations can be found 
as follows.   Define the tensors
\beq
{X^{ij}}_{kl}(u) ={\Delta^{ij}}_{kl} +\frac u4{\phi^{ij}}_{kl}
\eeq
where ${\Delta^{ij}}_{kl}=\frac 12(\delta_k^i\delta_l^j-\delta_l^i\delta_k^j)$
and the symbol ${\phi^{ij}}_{kl}\equiv \phi_{ijkl}$.
Then,  equations(\ref{feq})  can be written as follows
\beq
\Psi_{ijk}f_k = {X^{ij}}_{lm}(2)\{f_l,f_m\}
\eeq
Using now the algebra of the ${X^{ij}}_{kl}(u)$ tensors discussed in detail
in the Appendix we can prove that both the identities
\beq
\Psi_{ijk}{X^{jk}}_{lm}(-1) =0
\eeq
and
\beq
 {X^{ij}}_{mn}(-1){X^{mn}}_{kl}(2) = 0
\eeq
hold, and this terminates the second consistency check. 

%%%%%%%%%%% QUATERNIONS

\section{Octonionic and quaternionic formulation of the self-duality equations.}

The octonionic or Cauley alrgebra, is the appropriate structure to organize the seven
self duality equations\cite{gursey,fairlie}. The octonionic units $o_i$ satisfy
the algebra
\beq
o_i o_j = -\delta_{ij} + \Psi_{ijk} o_k
\label{omult}
\eeq
 where $i=1,\dots , 7$ are the 7 octonionic imaginary units with the property
\beq
\{o_i,o_j\} = - 2\delta_{ij}
\eeq
We choose the multiplication table\cite{gursey}
\beq
\Psi_{ijk}=\left\{\begin{array}{ccccccc}1&2&4&3&6&5&7\\
                             2&4&3&6&5&7&1\\
                             3&6&5&7&1&2&4
 \end{array}\right.
\label{2.1}
\eeq
In terms of these units an octonion  can be written as follows
\beq
{X} = x_0 o_0 + \sum_{i=1}^7 x_io_i
\eeq
with $o_0$ the identity element. The  conjugate is
\beq
\bar{X} = x_0 o_0 -  \sum_{i=1}^7 x_io_i
\eeq
The  octonions over the real numbers  can also be defined as pairs of quaternions
\beq
X = (x_1,x_2)\label{o-q}
\eeq
where $x_{1}= x_1^{\mu}\sigma_{\mu}$, $x_2 =x_2^{\mu}\sigma_{\mu}$ and the indices
$\mu $ run from $0$ to 3, while $x_{1,2}^0$ 
are real numbers and $x_{1,2}^i,i=1,2,3$ are imaginary numbers. Finally, 
$\sigma_{0}$ is  the Identity $2\times 2$ matrix and  $\sigma_{i}$ are the three
standard Pauli matrices.
\bea
\sigma_1  =\left(\begin{array}{cc}0&1\\
			            1&0
            \end{array}\right)
&
 \sigma_2 =\left(\begin{array}{cc}0&-\imath\\
                                   \imath&0
            \end{array}\right)
 &
\sigma_3 =\left(\begin{array}{cc}1&0\\
		                  0&-1
            \end{array}\right)
\eea
If $q = (q_1,q_2)$ and $r=(r_1,r_2)$ are two octonions,
the multiplication law is defined as
\beq
q*r \equiv (q_1,q_2)*(r_1,r_2) = (q_1r_1-\bar{r}_2r_2, r_2q_1+q_2\bar{r}_1)
\label{mr}
\eeq
where $q_1 = q_1^0+q_1^i\sigma_i$ and $\bar{q}_1= q_1^0-q_1^i\sigma_i$.
One can also define a conjugate operation for an octonion as
\beq
\bar{q} \equiv {\overline{(q_1,q_2)}} =(\bar{q}_1,-q_2)
\eeq
and we get the possibility to define  the norm and the scalar product
$q$~and~$r$
\bea
q\bar{q} & =& (q_1\bar{q}_1+\bar{q}_2q_2,0)\\
         &= & \sum_{\mu=0}^3\left({x_1^{\mu}}^2+{x_2^{\mu}}^2\right)\\
<q|r> &=& \frac 12 (q\bar{r}+\bar{q}r)
\eea   
In terms of the above formalism, the self duality equations can be written as follows
\beq
\dot{X}= \frac 12\{X,X\}\label{sdo}
\eeq
where now $X=X^io_i$ with $i=1,\cdots 7$ and the Poisson bracket for two octonions
is defined as
\beq
\{X,Y\} =\frac{\partial{X}}{\partial{\xi_1}}\frac{\partial{Y}}{\partial{\xi_2}}
         -\frac{\partial{X}}{\partial{\xi_2}}\frac{\partial{Y}}{\partial{\xi_1}}
\eeq
Using  now  (\ref{o-q}), and the 
multiplication rule (\ref{mr}) we can write eq(\ref{sdo}) as follows
\bea
\dot{x}_1 &=& \frac 12\left(\{x_1,x_1\}+\{\bar{x}_2,x_2\}\right)
\\
\dot{x}_2&=& -\{x_2,x_1\}\equiv \{x_2,\bar{x}_1\}
\eea
where $x_1 =  x_1^{\mu}\sigma_{\mu}$, $x_2=x_2^{\mu}\sigma_{\mu}$. 
Defining the octonionic units 
\bea
\begin{array}{cccc}
o_0=(1,0)&o_1=(\imath \sigma_1,0)&o_2=(\imath \sigma_2,0)&o_3=(-\imath \sigma_3,0)\\
o_4=(0,1)&o_5=(0,\imath \sigma_3)&o_6=(0,\imath \sigma_2),&o_7=(0,\imath \sigma_1)
\end{array}
\eea
we easily check, that the chosen multiplication table for the octonions
(\ref{2.1}) is satisfied and the seven coordinates 
 $X_i$ are grouped now as follows ($x_1^0 =0$)
\bea
x_1^i&=& \imath X_1,\imath X_2,\imath X_3\\
x_2^{\mu}&=& X_4,\imath X_7, \imath X_6,\imath X_5
\eea
and 
\bea
x_1& =&
 \left(\begin{array}{cc} X_3          & X_1-\imath X_2\\
                         X_1+\imath X_2& - X_3
        \end{array}\right)
\label{2.1a}\\
x_2& =&
\left(\begin{array}{rr} X_4+\imath X_5&X_6+\imath X_7\\
                      -(X_6-\imath X_7)&X_4-\imath X_5
     \end{array}\right)
\label{2.2}
\eea
The organization in (\ref{2.1a},\ref{2.2}) of the seven $X_i$ components obtained from
the quaternionic formulation will prove very useful to identify specific classes of
solutions as we will see in the next section.
\section{Embeddings of the three dimensional system.}

An obvious observation is that any three dimensional solution is also a solution of
the seven dimensional system discussed here. There are various ways however to
embed a three dimensional solution to the seven dimensional system. 
In this section, we discuss solutions of the self duality equations where the
coordinates $X_i$ are linear functions of the $SU(2)$ basis of functions on the 
sphere, which are the components of the unit vector in 3 dimensions written
in spherical coordinates\cite{FL}
\beq
 \{e_a,e_b\}= -\epsilon_{abc} e_c
\eeq
 Thus, our Ansatz is 
\beq
X_i(\xi_1,\xi_2,t)=A_i^a(t)e_a(\xi_1,\xi_2)
\label{XAe}
\eeq
and implies a generalised form of Nahm's equations 
\beq
\dot{A}_i^a =-\frac 12 \Psi_{ijk}A_j^bA_k^c e_{abc}, 
\label{eqA}
\eeq
where $a,b,c$ take the values $1,2,3$.
This Ansatz contains all the embeddings of the three dimensional system with $SU(2)$ 
symmetry which can be written explicitely as a $G_2$ rotation $R_{ij}$  of a seven-vector with
non zero components  the first three.
\beq
A_i^a = R_{ij}B_j^a
\eeq
where $B_j^a$ is defined through the three dimensional $SU(2)$ solution 
\beq
B_i^a = (T_1^a,T_2^a,T_3^a,0,0,0,0)
\eeq
Here the matrix $T_a^b$ , $a,b=1,2,3$ satisfies the three dimensional Nahm equations.

Let us now present some simple cases: The grouping of coordinates in relations
(\ref{2.1a},\ref{2.2})  is suggestive  in order to write the self duality 
 equations in terms
of the complex coordinates $X_{\pm} =X_1\pm \imath X_2,$,  $Y_{\pm} =X_4\pm \imath X_5,$
and  $Z_{\pm} =X_6\pm \imath X_7$. In terms of the later
the system can be written as follows
\bea
\dot{X}_+ &=&\imath (\{X_3,X_+\}+\{Y_+,Z_-\})\\
\dot{Y}_+&=& \imath(\{Y_+,X_3\}+\{X_+,Z_+\})\\
\dot{Z}_+&=&\imath ( \{X_3,Z_+\}+\{X_-,Y_+\})\\
\dot{X}_3&=&\imath \frac 12(\{X_+,X_-\}+\{Z_+,Z_-\}-\{Y_+,Y_-\})
\eea
We can easily obtain
 some simple solutions of the system in five or seven
dimensions. In five dimensions in particular, we set $X_+=\imath Y_-$ and
 $Z_+=0$. Then,
we find that the system is reduced in the three dimensional case\cite{FL}
with the identifications
\beq
X_{\pm} \ra A_{\pm}/\sqrt{2}\;\,\, , \;\,\;\,\, X_3\ra A_3
\eeq
Another solution which  embeds every  solution of the 3 dimensional
case in seven dimensions, can be obtained by the
 identifications
$X_+ =Z_+ = \imath Y_-$. This  solution is reduced to that of the three 
dimensional
one\cite{FL} with the following rescaling
\beq
X_{\pm} \ra A_{\pm}/\sqrt{3}\;\,\, , \;\,\;\,\, X_3\ra A_3
\eeq
%so all the solutions of the three dimensional case 
%\cite{FL} are automatically solutions of the
%five and seven dimensional equations. 
 An explicit construction
shows that the two solutions are connected with 
the  orthogonal 
transformation $|\xi_7> = {\cal O}|\xi_3>$ where the matrix ${\cal O}$ is
\beq
{\cal O} =
 \left(\begin{array}{ccccccc}a&0&0&-b&0&b&a \\
                             0&a&0&0&-b&a&-b\\
                             0&0&1&0&0&0&0  \\
			     0&a&0&0&-b&-a&b\\
			     a&0&0&-b&0&-b&-a\\
			     a&0&0&2b&0&0&0 \\
  			     0&a&0&0&2b&0&0
        \end{array}\right)
\label{2.3}
\eeq
where $a=\frac 1{\sqrt{3}}$ and $b=\frac 1{\sqrt{6}}$, $<\xi_3|=(X_1,X_2,X_3,0,0,0,0)$ and
$|\xi_7>$ stands for the seven dimensional vector. 

We conclude this work by summarizing our results.
The relation of the octonionic algebra with quaternions gives a useful formulation
of the self - duality equations which extends in a natural way the three
dimensional system and the corresponding  generalized Nahm's equations for $SDiff S_2$.
By intoducing in the place of $SU(2)$ algebra of functions on the sphere, a quadratic
algebra of seven functions with $G_2$ symmetry, we succeeded to factorize the time dependence
in a simple way which may facilitate the study of solutions of the self duality equations.
Although the general system of self duality equations in seven dimensions does not seem
to have a Lax pair, at least in a direct way, due to the non-associativity of the
octonionic algebra, it may happen that there is a generalization of the zero-curvature 
condition under which this system is integrable. In the case of three dimensions
the restriction of the solutions to the $SU(2)$ subalgebra of functions on the 
spherical membrane reduces the problem to the study of Nahm's SU(2) equations.
In the same way, in seven dimensions  the introduction of the quadratic algebra of
functions on the sphere reduces the problem to the generalization of Nahm's 
equations  with similar scaling properties with respect to time. This gives indications
that the specific system maybe relevant for the study of monopole type of configurations
of membranes.
 
The relevance of the self duality membrane equations in seven dimensions for the spectrum
of instantons of the 11-dimensional supermembrane is an open problem as well as the
number of supersymmetries surviving  these solutions.

\vspace*{1cm}

{{\bf Acknowledgements:}{\it \, We would like to thank CERN theory division
for kind hospitality where
 this work has been done.One of us (E.G.F.) would like to thank D. Fairlie for
initiating  him into the field of octonionic duality.
 We also thank I. Antoniadis, I. Bakas and E. Kiritsis for
useful discussions.}

\newpage
\section{Appendix}
In this Appendix we derive the properties
 of the tensors ${X^{ij}}_{kl}(u)$ used 
in section 3 to make consistency checks of our Ansatz. 
Consider the generalized matrices ${{\cal P}^{ij}}_{kl}(u,v)$. 
\beq
{{\cal P}^{ij}}_{kl}(u,v) = u{\Delta^{ij}}_{kl}+\frac v4{\phi^{ij}}_{kl}
\eeq
Using the properties 
\bea
{\Delta^{ij}}_{kl}{\Delta^{mn}}_{ij}& =& {\Delta^{mn}}_{kl}\\
{\phi^{ij}}_{mn}{\phi^{mn}}_{kl} &=& 8 \left({ \Delta^{ij}}_{kl}
 +\frac 14{\phi^{ij}}_{kl}\right)\\
{\Delta^{ij}}_{kl}{\phi^{kl}}_{mn}&=& {\phi^{ij}}_{mn}
\eea
we  derive the following multiplication rule
\beq
{{\cal P}^{ij}}_{kl}(u_1,v_1){{\cal P}^{ij}}_{kl}(u_2,v_2)
={{\cal P}^{ij}}_{kl}(u_3,v_3)
\eeq
where
\bea
u_3&=& u_1u_2+\frac{v_1v_2}2\nonumber\\
v_3&=& u_1v_2+u_2v_1+\frac{v_1v_2}2\nonumber
\eea
We observe that this is a group structure which can be realised as a subgroup
of the general linear group in two dimensions through the matrices
\beq
G({u, v})=\left(\begin{array}{cc}u& v/2\\
                             v&u+ v/2
 \end{array}\right)
\eeq
{}For the existence of the inverse, one should restrict the parameters $u,v$
inside the angular regions 
\bea
v & =&  u\\
v & = & - 2u 
\eea
{}For the case of $u=1$, we restrict to ${{\cal P}^{ij}}_{kl}(1,v)
\equiv {X^{ij}}_{kl}(v)$ matrices. 
Using the above, we find the multiplication law
\bea
{X^{ij}}_{kl}(u){X^{mn}}_{ij}(v)& =& 
(1+\frac{uv}2) {\Delta^{ij}}_{kl}+\frac 14(u+v+\frac{uv}2)
{\phi^{ij}}_{kl}\nonumber\\
&\equiv&
 (1+\frac{uv}2) {X^{mn}}_{kl}(w)\label{alg}
\eea
where
\beq
w=\frac{u+v+uv/2}{1+{uv}/2}\label{w}
\eeq
{}Form the properties of the symbol $\Psi_{ijk}$, we find
\beq
\Psi_{ijk}{X^{jk}}_{lm}(u) = (1+u) \Psi_{ilm}
\eeq
so for $u=-1$, the ${X}^{jk}(u)$ antisymmetric matrices satisfy the costraint
for the $G_2$ algebra
\beq
\Psi_{ijk}{X^{jk}}_{lm}(-1) =0
\eeq
 
We finally observe the following interesting projective properties for the
end points of the group parameter $u$
\bea
{X^{ij}}_{mn}(u){X^{mn}}_{kl}(2)& =& (1+u) {X^{ij}}_{kl}(2)\\
{X^{ij}}_{mn}(u){X^{mn}}_{kl}(-1)& =& (1-\frac u2) {X^{ij}}_{kl}(-1)
\eea

\end{document}